\documentclass{article}
\usepackage[utf8]{inputenc}
\usepackage{spconf,amsmath,graphicx}
\usepackage{xcolor}
\graphicspath{{./}}
\usepackage{subfigure}
\usepackage[T1]{fontenc}
\usepackage{csquotes}
\usepackage{float}

\name{Qing Zou, Abdul Haseeb Ahmed, Prashant Nagpal, Stanley Kruger, Mathews Jacob.}
\address{University of Iowa}
\begin{document}

\title{Deep Generative SToRM model for dynamic imaging}

\maketitle

\begin{abstract}
We introduce a novel generative smoothness regularization on manifolds (SToRM) model for the recovery of dynamic image data from highly undersampled measurements. The proposed generative framework represents the image time series as a smooth non-linear function of low-dimensional latent vectors that capture the cardiac and respiratory phases. The non-linear function is represented using a deep convolutional neural network (CNN). Unlike the popular CNN approaches that require extensive fully-sampled training data that is not available in this setting, the parameters of the CNN generator as well as the latent vectors are jointly estimated from the undersampled measurements using stochastic gradient descent. We penalize the norm of the gradient of the generator to encourage the learning of a smooth surface/manifold, while temporal gradients of the latent vectors are penalized to encourage the time series to be smooth. The main benefits of the proposed scheme are (a) the quite significant reduction in memory demand compared to the analysis based SToRM model, and (b) the spatial regularization brought in by the CNN model. We also introduce efficient progressive approaches to minimize the computational complexity of the algorithm.
\end{abstract}

\section{Introduction}

The quest for high spatial and temporal resolution is central to several dynamic imaging problems, ranging from MRI, video imaging, to microscopy. 
A popular approach to improve spatio-temporal resolution is self-gating, where cardiac and respiratory information is estimated from navigator or central k-space using bandpass filtering or clustering, followed by binning and reconstruction \cite{feng2014golden,christodoulou2018magnetic}.
Several authors have also introduced smooth manifold regularization, which models the images in the time series as points on a high dimensional manifold \cite{poddar2015dynamic,poddar2019manifold,nakarmi2017kernel}.
 This approach may be viewed as an implicit soft-gating alternative to self-gating methods. Manifold methods including our smoothness regularization on manifolds (SToRM) approach has been demonstrated in a variety of dynamic imaging applications with good performance \cite{poddar2015dynamic,poddar2019manifold,nakarmi2017kernel}.
 Since the data is not explicitly binned into a specific phase, manifold methods are not vulnerable to potential errors in clustering the time series based on navigators. Despite the benefits, a key challenge with current manifold methods is the high memory demand. Unlike self-gating methods that only recover the specific phases, manifold schemes recover the entire time series. This approach restricts the extension of the framework to higher dimensional problems. The high memory demand also makes it difficult to use additional spatial and temporal regularization.

The main focus of this work is to exploit the power of deep convolutional neural networks (CNN) to introduce an improved and memory efficient generative/synthesis formulation of SToRM. Unlike current manifold and self-gating methods, this approach does not require k-space navigators to estimate the motion states. Besides, unlike traditional CNN based approaches, the proposed scheme does not require extensive training data, which is challenging to acquire in free-breathing applications. We note that current manifold methods can be viewed as an analysis formulation. Specifically, a non-linear injective mapping is applied on the images such that the mapped points of the alias-free images lie on a low-dimensional subspace. When recovering from undersampled data, the nuclear norm prior is applied in the transform domain to encourage their non-linear mappings to lie in a subspace. Unfortunately, this analysis approach requires the storage of all the image frames in the time series. In this work, we model the images in the time series as non-linear mappings \(\mathbf \rho_t= \mathcal G_{\theta}\left(\mathbf z_t\right)\), where \(\mathbf z_i\) are vectors that live in a very low-dimensional subspace. The dimension of the subspace can be very small (e.g 2-4) in practical applications. We represent the non-linear mapping using a convolution neural network with weights \(\theta\). The memory footprint of the algorithm depends on the number of parameters \(\theta\) and \(z\), which is orders of magnitude smaller than that of traditional manifold methods.

\begin{figure}[t!]
	\centering
	\includegraphics[width=0.4\textwidth]{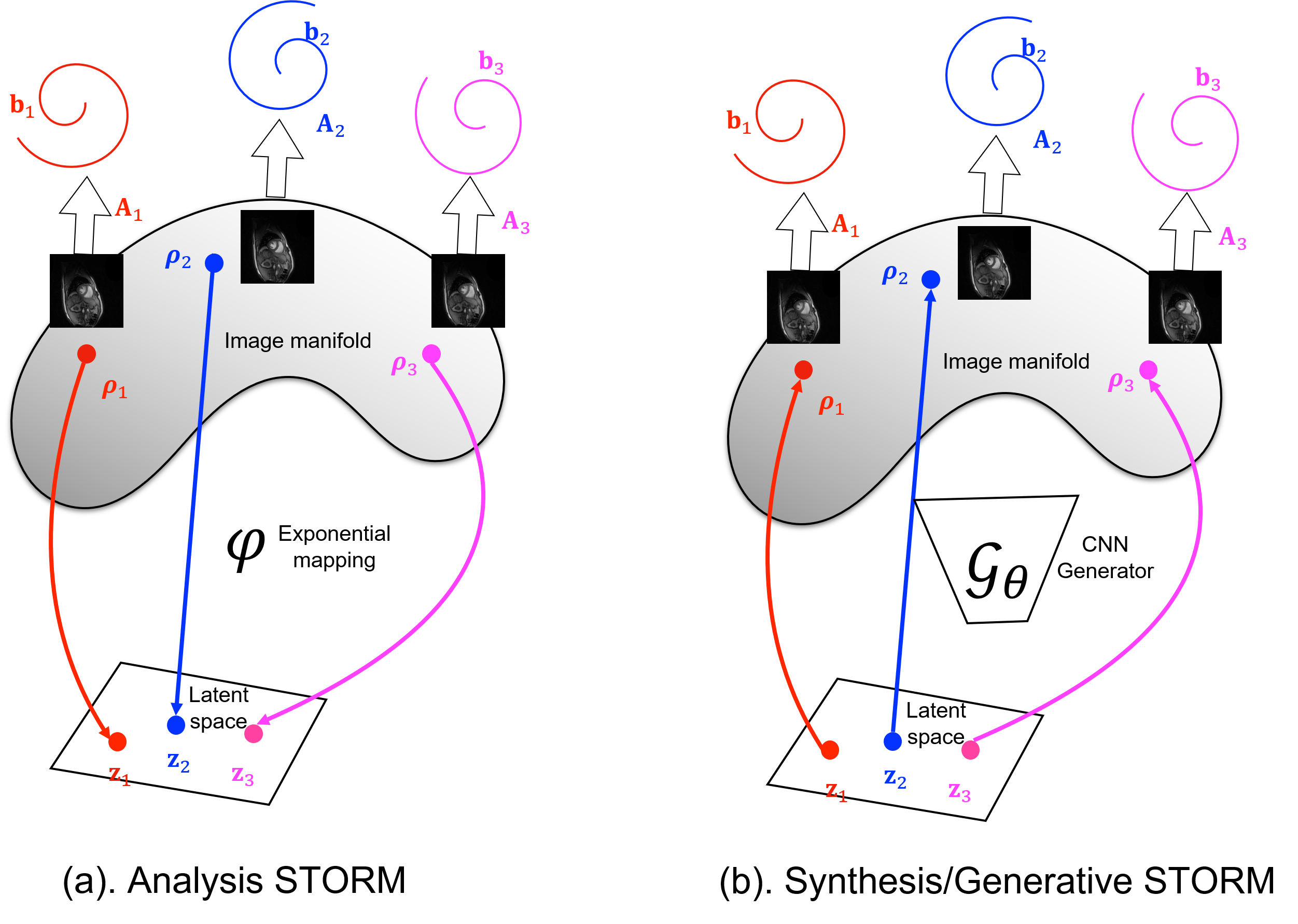}\vspace{-1.5em}
	\caption{\small (a) Analysis SToRM and (b) Generative SToRM. The analysis formulation \cite{poddar2019manifold,ahmed2020free} in (a) minimizes the nuclear norm of the non-linear mappings \(\varphi(\mathbf x_i)\) of the images \(\mathbf x_i\) to encourage them to be in a subspace. By contrast, the proposed formulation expresses the images as non-linear mappings \(\mathcal G_{\theta}(\mathbf z_i)\) of the low-dimensional latent vectors \(\mathbf z_i\). The main benefit of the generative model is its ability to compress the data, thus offering a memory efficient algorithm. }\vspace{-1em}
	\label{stormmethods}
\end{figure}

We propose to jointly optimize for the network parameters \(\theta\) and the latent vector \(\mathbf z\) such that the cost \(\sum_i\|\mathcal A_t(\mathcal G_{\theta}\mathbf z_t) - \mathbf b_i\|^2\) is minimized during image reconstruction. The smoothness of the manifold generated by \(\mathcal{G}_{\theta}(\mathbf z)\) depends on the gradient of \(\mathcal G_{\theta}\) with respect to its input. To obtain a smooth manifold, we regularize the gradient of the mapping \(\|\nabla_{z} \mathcal G_{\theta}\|^2\). 
Similarly, the images in the time series are expected to vary smoothly in time. Hence, we also use a Tikhonov smoothness penalty on the latent vectors \(\mathbf z_t\) to further constrain the solutions. Unlike traditional CNN methods that are fast during testing/inference, the direct application of this scheme to the dynamic MRI setting is computationally expensive. We use a three-step progressive-in-time approach to significantly reduce the computational complexity of the algorithm. Specifically, we grow the number of frames in the datasets during the optimization process. 
The latent vectors from the previous iteration are linearly interpolated to initialize the latent vectors. 
We observe that the use of the progressive-in-time approach significantly reduces the computational complexity of the algorithm.

The proposed approach is inspired by deep image prior (DIP) \cite{ulyanov2018deep}, which was introduced for static imaging problems. We note that the extension of DIP to dynamic imaging was considered in \cite{jin2019time}. The key difference of the proposed formulation from the above work is the joint optimization of the latent variables \(\mathbf z\), unlike the above method that chooses \(\mathbf z\) as random or interpolated versions of random vectors. Another key distinction is the use of regularization priors on the network parameters and latent vectors, which ensures that the scheme learns meaningful latent vectors and the performance of the network does not degrade with iterations as in traditional DIP methods. 

\section{Methods}

Smooth manifold methods model images \(\mathbf x_i\) in the dynamic time series as points on a smooth manifold. In SToRM, the exponential (injective) functions of the images denoted by \(\varphi(\mathbf x_i)\) of the alias-free images are assumed to lie on a low-dimensional subspace. See Fig. \ref{stormmethods}.(a).
The joint recovery of the images denoted by the matrix \(\mathbf X = \left[\mathbf x_1,..\mathbf x_N\right]\) from undersampled data is posed as a nuclear norm minimization problem
\begin{equation}\label{storm}
\mathbf X^* = \arg \min_{\mathbf X} \|\mathcal A(\mathbf X) - \mathbf B\|^2 + \lambda~ \|\left[\varphi(\mathbf x_1),..,\varphi\_\{t\}(\mathbf x_N)\right]\|_*
\end{equation}

To overcome the challenges with the above analysis scheme, we propose to model the images in the time series as
\begin{equation}\label{genmodel}
\mathbf x_i = \mathcal G_{\theta}(\mathbf z_i),
\end{equation}
where \(\mathcal G_{\theta}\) is a non-linear mapping. We realize \(\mathcal G_{\theta}\) using a deep convolutional neural network, inspired by the extensive work on generative image models.
Here, \(\mathbf z_i\) are latent vectors that lie in a low-dimensional subspace. As \(\mathbf z_i\) vary in the subspace, their non-linear mappings vary on the image manifold. The mapping \(\mathcal G_{\theta}\) may be viewed as the inverse of the injective mapping \(\varphi\) considered in analysis SToRM; rather than mapping the images to a low-dimensional subspace as in classical SToRM methods we now propose to express the images as non-linear functions of latent variables living in a low-dimensional subspace. See Fig. {\ref{stormmethods}.(b).

The smoothness of the manifold is determined by the gradient of the non-linear mapping, denoted by \(\nabla_{\mathbf z} \mathcal G_{\theta}\). A mapping with high gradient values can result in very similar latent vectors being mapped to very different images. To minimize this risk, we propose to penalize the \(\ell_2\) norm of the gradients of the network, denoted by \(\|\nabla_{\mathbf z} \mathcal G_{\theta}\|^2\). We term this prior as network regularizer. We expect the adjacent time frames in the time series to be similar; we propose to add a temporal smoothness regularizer on the latent vectors. The parameters of the network \(\theta\) as well as the low-dimensional latent vector \(\mathbf z\) are estimated from the measured data by minimizing

\vspace{-2em}
\begin{eqnarray}\nonumber\label{cost}
\mathcal C(\mathbf z,\theta)&=& \sum_{i=1}^N\|\mathcal A_i\left(\mathcal G_{\theta}[\mathbf z_i]\right) - \mathbf b\|^2 + \lambda_1 \underbrace{\|\nabla_{\mathbf z} \mathcal G_{\theta}\|^2}_{\scriptsize \mbox{network regularization}}\\\label{gen-storm}&&\qquad + \lambda_2 \underbrace{\|\nabla_{t} \mathbf z_t\|^2 }_{\scriptsize\mbox{temporal regularization}}
\end{eqnarray}
with respect to \(\mathbf z\) and \(\theta\). We initialize the network parameters and latent vectors to be random variables.

We use ADAM optimization to determine the optimal parameters. Note that the first and the second term in the expression is separable over \(i\). To keep memory demand of the algorithm low, we propose to choose mini-batches consisting of random subset of frames. A key benefit of this framework over conventional neural network schemes is that it does not require any training data. Note that it is often impossible to acquire fully-sampled training data in dynamic imaging applications.

The main benefit of this model is the compression offered by the representation; the number of parameters of the model in \eqref{genmodel} is orders of magnitude smaller than the number of pixels in the dataset. The dramatic compression offered by the representation, together with the mini-batch training provides a memory efficient alternative to analysis SToRM \cite{poddar2015dynamic,poddar2019manifold}. Although our focus is on establishing the utility of the scheme in 2-D settings in this paper, the approach can be readily translated to higher dimensional applications. Another benefit is the implicit spatial regularization brought in by the generative CNN. Specifically, CNNs are ideally suited to represent images rather than noise-like alias artifacts \cite{ulyanov2018deep}.

\begin{figure}[t]
	\centering
	\includegraphics[width=0.25\textwidth]{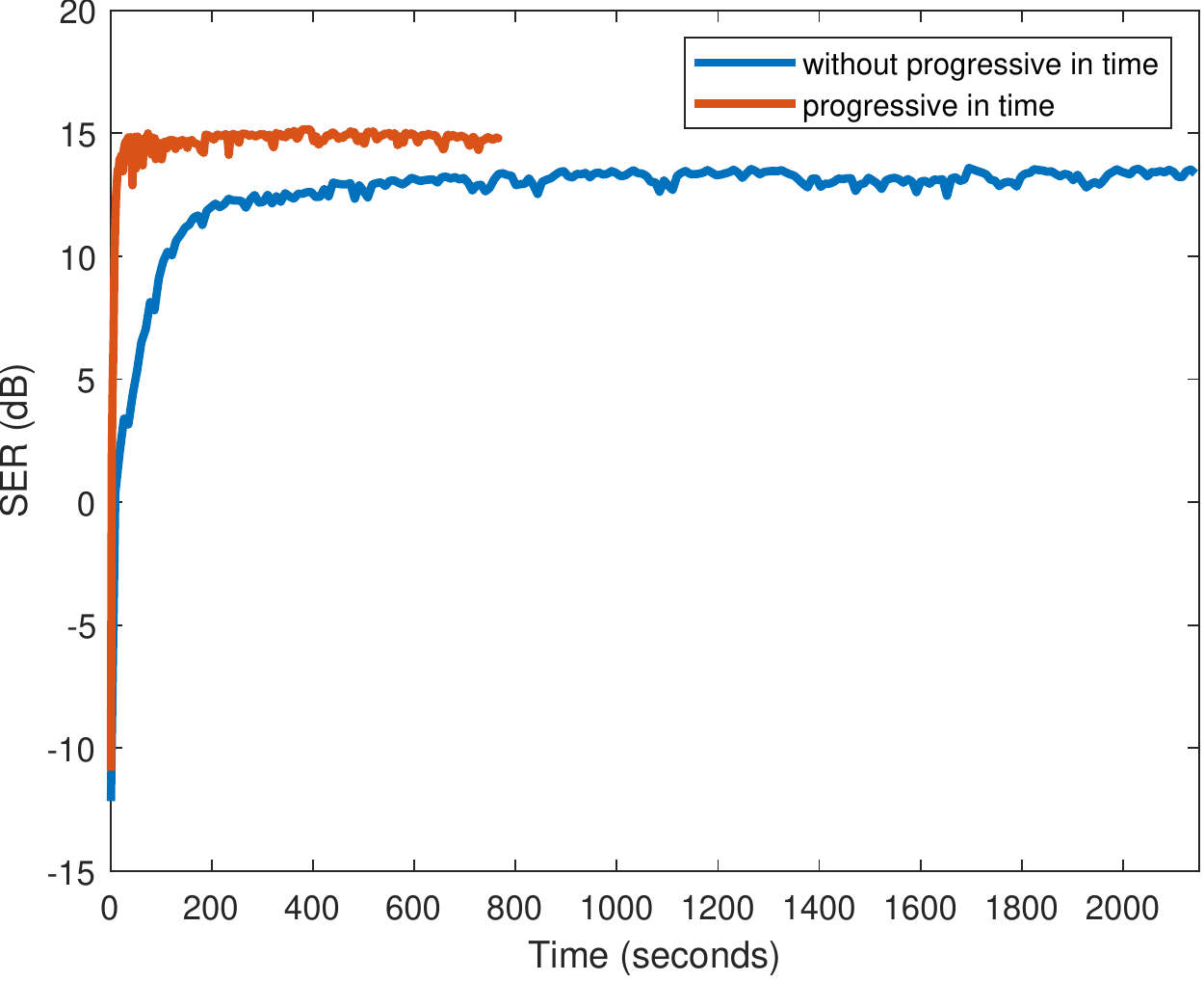}\vspace{-1em}
	\caption{\small Reconstruction performance with progressive training in time and without progressive training in time. From the plot, one can see that progressive training in time produces better results with much less running time comparing to the training without progressive in time.}\vspace{-1em}
	\label{PGvsNPG}
\end{figure}

\begin{figure}[b!]
\vspace{-1em}
	\centering
	\includegraphics[width=0.23\textwidth]{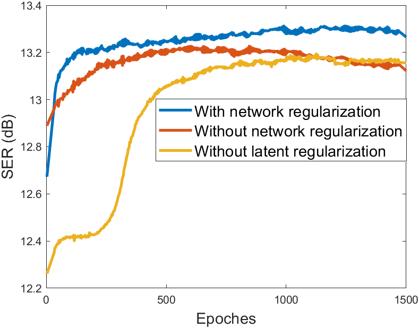}\quad
	\includegraphics[width=0.23\textwidth]{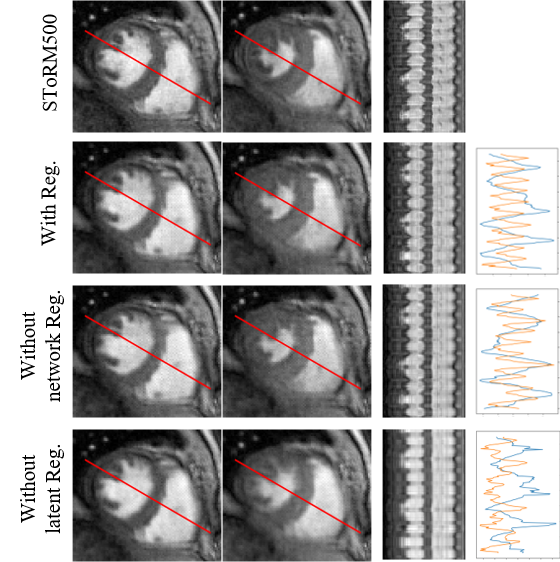}\vspace{-1em}
	\caption{\small Impact of network regularization and latent variable regularization. The SER vs epoch plots are shown above, while two of the reconstructed images, their time profiles, and recovered latent variables are shown. We note that the blue curve captures respiratory motion, while the orange one captures cardiac motion.}
	\label{RegvsNReg}\vspace{-1em}
\end{figure}

\vspace{-1.5em}
\subsection{Progressive in time training}
\vspace{-0.5em}

\label{pit}
While the generative SToRM approach significantly reduces the memory demand, a challenge with this approach is the increased computational complexity. To minimize the complexity, we propose to use a progressive optimization strategy. Specifically, we solve for a sequence of vectors \(\mathbf z_0\), \(\mathbf z_1\),.., \(\mathbf z_M\) each corresponding to increasing number of time frames. For instance, in this work we choose \(\mathbf z_0\) to be a \(2\times 1\) vector, where we consider the recovery of an average image \(\mathcal G_{\theta}(\mathbf z_0)= \mathbf x_0\) from the entire data. We solve for the optimal \(\theta_0\) and \(\mathbf z_0\) by minimizing \eqref{storm}. Since we are solving for a single image, this optimization is fast. Following convergence, the latent vector \(\{\mathbf z_0\}\) is linearly interpolated to the size of \(\mathbf z_1\) and used along with \(\theta_0\) as initialization, while solving for \(\left\{\theta_1,\mathbf z_1\right\}\). This approach significantly reduces the computational complexity as seen from our experiments

\begin{figure}[b!]
\vspace{-1em}
	\centering
	\includegraphics[width=0.45\textwidth]{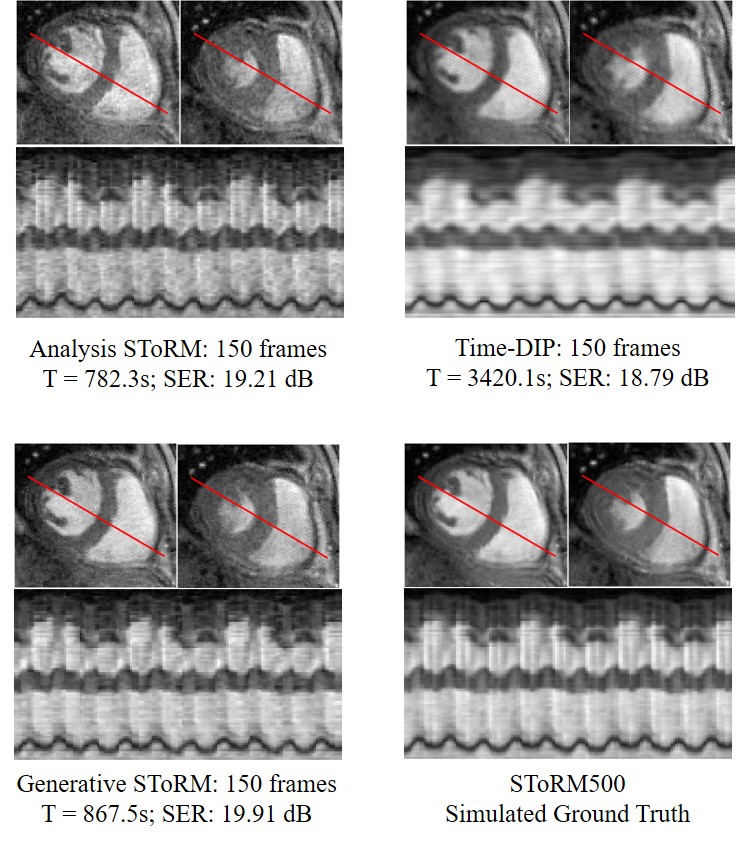}\vspace{-1em}
	\caption{Comparison of Generative SToRM, Analysis SToRM, time dependent deep image prior.}
	\label{result_comparison}\vspace{-1em}
\end{figure}

\vspace{-0.5em}
\section{Experiments}
\vspace{-0.5em}
\subsection{Dataset and imaging experiments}

All the experiments in this paper are based on a whole-heart multi-slice dataset collected in the free-breathing mode using a golden angle spiral trajectory. The acquisition of the data was performed on a GE 3T scanner. The sequence parameters were: TR= 8.4 ms, FOV= 320 mm x 320 mm, flip angle= 18 degrees, slice thickness= 8 mm.

Results were generated using an Intel Xeon CPU at 2.40 GHz and a Tesla P100-PCIE 16GB GPU. Results in \S 4.2, \S 4.3 were based on the first slice in the dataset, and results in \S 4.4, \S 4.5 were based on the second slice in the dataset. We binned the data from six spiral interleaves corresponding to 50 ms temporal resolution. The entire dataset corresponds to 522 frames. We omit the first 22 frames and used the remaining 500 frames for SToRM reconstructions, which is used as ground truth for comparisons. In all the studies, we assumed the latent variables to be two dimensional since the main source of variability in the data correspond to cardiac and respiratory motion.
\vspace{-1em}
\subsection{Benefit of progressive in time approach}

We demonstrate the quite significant reduction in running time offered by the progressive training strategy described in Section \ref{pit} in Fig. \ref{PGvsNPG}. Here, we consider the recovery from 150 frames with and without the progressive strategy. We plot the reconstruction performance, measured by the Signal-to-Error Ratio (SER) with respect to the running time. The results show that the proposed scheme can offer good reconstructions in \(\approx 200\) seconds, which is better than the direct approach that takes more than 2000 seconds.
 \vspace{-1em}
\subsection{Impact of regularization priors}
We study the impact of network regularization priors in Fig. \ref{RegvsNReg}.(a), where we show the reconstruction performance with respect to the number of epochs. The recovered latent variables are also shown in the plots. We chose \(\lambda_2=2\) in this experiment. We note that unlike the case without network regularization, the SER of the regularized reconstruction increases with iteration. The case without regularization will start to fit to the noise with iterations as in the case of deep image prior. We note that with regularization, the latent variables capture cardiac (orange curve) and respiratory (orange curve) motion, even though no explicit priors or additional information (e.g navigators) about cardiac or respiratory rates were used. Without network regularization, we observe increased mixing of the cardiac and respiratory patterns in the latent vectors.

In the cost function \eqref{cost}, we also have the temporal smoothness regularization of the latent variables. We compare \(\lambda_2=2\) against \(\lambda_2=0\), while \(\lambda_1\) was fixed as \(0.001\). Similar to the network regularization setting, we observe that the performance of the un-regularized algorithm falls with iterations, while the performance of the regularized approach increases or plateau with iterations. We also obsrved significant mixing between cardiac and respiratory patterns in the latent variables when no regularization is used.

\vspace{-1em}
\subsection{Comparison with existing methods}
We compare the proposed generative SToRM approach with analysis SToRM \cite{ahmed2020free} and time dependent deep image prior algorithm \cite{jin2019time}. We use the k-space data of 150 frames for the reconstructions. The reconstruction results are shown in Fig. \ref{result_comparison}. The results show that the generative SToRM approach is able to reduce noise and alias artifacts compared to analysis SToRM, offering around 1dB improvement in performance. We attribute the improved performance to spatial regularization offered by the CNN generator, which is absent in the analysis SToRM formulation. The reconstruction time of both the algorithms are comparable. The Time-DIP scheme, which assumes the latent variables to be fixed as random values results in increased artifacts and blurring of motion details. We note that unlike the analysis schemes, the proposed scheme does not use k-space navigators to estimate the motion states; the latent variables are estimated from the measured k-space data itself.

\vspace{-1em}
\section{Conclusion}
\vspace{-1em}

We introduce a generative manifold representation for the recovery of dynamic image data from highly undersampled measurements. The deep CNN generator is used to lift low-dimensional latent vectors to the smooth image manifold and this proposed scheme does not require fully-sampled training data. We jointly optimize the CNN generator parameters and the latent vectors based on the undersampled data. We also proposed the training-in-time approch to minimize the computational complexity of the algorithm. During the training, the norm of the gradients of the generator is penalized to the learning of a smooth surface/manifold, while temporal gradients of the latent vectors are penalized to encourage the time series to be smooth. Comparisons with existing methods suggest the utility of the proposed scheme in dynamic images.

\vspace{-1em}
\section{Compliance with Ethical Standards}
\vspace{-1em}

This research study was conducted using human subject data. The institutional review board at the local institution approved the acquisition of the data, and written consent was obtained from the subject.

\vspace{-1em}
\section{Acknowledgments}
\vspace{-1em}

This work is supported by grants NIH 1R01EB019961-01A1 and R01EB019961-02S. The authors claim that there is no conflicts of interest.




\vspace{-1em}
\bibliographystyle{IEEEbib}
\bibliography{refs}

\begin{thebibliography}{1}

\bibitem{feng2014golden}
Li~Feng, Robert Grimm, Kai~Tobias Block, Hersh Chandarana, Sungheon Kim, Jian
  Xu, Leon Axel, Daniel~K Sodickson, and Ricardo Otazo,
\newblock ``Golden-angle radial sparse parallel mri: combination of compressed
  sensing, parallel imaging, and golden-angle radial sampling for fast and
  flexible dynamic volumetric mri,''
\newblock {\em Magnetic resonance in medicine}, vol. 72, no. 3, pp. 707--717,
  2014.

\bibitem{christodoulou2018magnetic}
Anthony~G Christodoulou, Jaime~L Shaw, Christopher Nguyen, Qi~Yang, Yibin Xie,
  Nan Wang, and Debiao Li,
\newblock ``Magnetic resonance multitasking for motion-resolved quantitative
  cardiovascular imaging,''
\newblock {\em Nature biomedical engineering}, vol. 2, no. 4, pp. 215--226,
  2018.

\bibitem{poddar2015dynamic}
Sunrita Poddar and Mathews Jacob,
\newblock ``Dynamic mri using smoothness regularization on manifolds (storm),''
\newblock {\em IEEE transactions on medical imaging}, vol. 35, no. 4, pp.
  1106--1115, 2015.

\bibitem{poddar2019manifold}
Sunrita Poddar, Yasir~Q Mohsin, Deidra Ansah, Bijoy Thattaliyath, Ravi Ashwath,
  and Mathews Jacob,
\newblock ``Manifold recovery using kernel low-rank regularization: Application
  to dynamic imaging,''
\newblock {\em IEEE Transactions on Computational Imaging}, vol. 5, no. 3, pp.
  478--491, 2019.

\bibitem{nakarmi2017kernel}
Ukash Nakarmi, Yanhua Wang, Jingyuan Lyu, Dong Liang, and Leslie Ying,
\newblock ``A kernel-based low-rank (klr) model for low-dimensional manifold
  recovery in highly accelerated dynamic mri,''
\newblock {\em IEEE transactions on medical imaging}, vol. 36, no. 11, pp.
  2297--2307, 2017.

\bibitem{ahmed2020free}
Abdul~Haseeb Ahmed, Ruixi Zhou, Yang Yang, Prashant Nagpal, Michael Salerno,
  and Mathews Jacob,
\newblock ``Free-breathing and ungated dynamic mri using navigator-less spiral
  storm,''
\newblock {\em IEEE Transactions on Medical Imaging}, 2020.

\bibitem{ulyanov2018deep}
Dmitry Ulyanov, Andrea Vedaldi, and Victor Lempitsky,
\newblock ``Deep image prior,''
\newblock in {\em Proceedings of the IEEE Conference on Computer Vision and
  Pattern Recognition}, 2018, pp. 9446--9454.

\bibitem{jin2019time}
Kyong~Hwan Jin, Harshit Gupta, Jerome Yerly, Matthias Stuber, and Michael
  Unser,
\newblock ``Time-dependent deep image prior for dynamic mri,''
\newblock {\em arXiv preprint arXiv:1910.01684}, 2019.

\end{thebibliography}
\vspace{-1em}
\end{document}